# Research on Sectionalizing Switches Placement Problem of Distribution System Automation Based on Multi-Objective Optimization Analysis

Selma Cheshmeh Khavar
Dept. of Electrical Engineering
Amirkabir University of Technology
Tehran, Iran
s_cr@aut.ac.ir

Arya Abdollahi
Dept. of Electrical and Information Engineering
Polytechnic University of Bari
Bari, Italy
arya.abdollahi@poliba.it

*Abstract* - Achieving high distribution-reliability levels and concurrently minimizing operating costs can be considered as the main issues in distribution system optimization. Determination of the optimal number and location of automation devices in the distribution system network is an essential issue from the reliability and economical points of view. To address these issues, this paper develops a multi-objective model, wherein the primary objective, optimal automation devices placement is implemented aiming at minimizing the operating costs, while in the second objective the reliability indices improvement is taken into account. So, modified non dominated sorting genetic algorithm, is developed and presented to solve this multi-objective mixed-integer non-linear programming problem. The feasibility of the proposed algorithm examined by application to two distribution feeders of the Tabriz distribution network containing the third feeder of the Azar substation with a distributed generation unit and first and third feeders of ElGoli substation which form a double feed feeder.

## Nomenclature

| | |
|---|---|
| $F_1$ | Switch placement cost function |
| $DS_i$ | Decision variable related to the installation or non-installation of switch devices at the candidate site |
| $CS$ | Construction cost of a switch device |
| $n$ | Number of candidate locations for switch installation |
| $DT_j$ | Decision variable related to the construction or non-construction of maneuver point at the candidate site $j$ |
| $CT_j$ | Minimum cost of the maneuver point construction on the candidate site $j$ |
| $m$ | Number of candidate locations for maneuver point construction |
| $MC$ | Cost of equipment maintenance and operation |
| $P_W$ | Economic factor to convert current costs of the study horizon to the current costs |
| $Infr$ | Annual inflation rate |
| $Intr$ | Annual interest rate |
| $ny$ | Study horizon by year |
| $ENS_T$ | Total energy not supply |
| $ENS_i$ | Energy not supply in the distribution post $i$ |
| $IC_i$ | Average cost of blackout in the distribution post $i$ |
| $K_i$ | Demand importance factor of $i$th post |
| $nLP$ | Number of distribution points on the feeder |
| $AIC_i$ | Blackout cost |
| $P_i$ | The percentage of each load in distribution post $i$ |
| $\lambda_i$ | Failure rate in $i$th mode |
| $r_i$ | Repair rate |
| $P$ | Average consumption at the load point |
| $\eta_c$ | Crossover index |
| $\eta_m$ | Mutation index |
| $V_i$ | Voltage of the $i$th load point |
| $I_i$ | Current of the $i$th load point |
| $f_{i-1}^k$ | $k$th objective of the ($i$-$1$)th individual |
| $f_{i+1}^k$ | $k$th objective of the ($i$+$1$)th individual |
| $r$ | Iteration number |

## I. Introduction

The issues and challenges in the distribution network are currently one of the most pressing concerns in the energy area, attracting the attention of all operation professionals. High power losses, voltage drops, and power interruptions are these problems, for which the installation of distribution automation devices is a crucial solution, given the enormous amount of investment and the requirement for optimal operation in this type of network. All of the end-consumers desire to get high-quality power with no disruptions. As a result, the distribution network's functioning is guided by the following two key ideas. 1) Consistency in delivering customer service 2) Keeping a high level of service quality. Due to the expanse of Tabriz city distribution network, the requirement for optimal placement of distribution network automation equipment is becoming increasingly apparent. The goal of distribution system automation is to reduce the amount of total power loss, energy not supply (ENS), the staff salaries. Aside from the technical benefits, the automation of distribution networks has economic profitability, which is dependent on the optimum selection of the function.

A reactive tabu search optimization technique is used in [1] to find the best locations for automation devices. The proposed strategy is considered the cost of long-term interruptions, automation equipment's purchase, and maintenance in the objective function. The author in [2] is studied a novel objective containing the cost of long-term interruptions and automation of distribution systems. To solve the suggested mixed-integer non-linear programming problem, the author is proposed a modified particle swarm optimization technique. The heuristic combinatory search technique developed in [3] decomposes the overall automation issue involving numerous types of automation devices into a series of simple sub-problems involving a single type of device. It specifies a set of heuristic criteria for determining the position of single feeding devices. The expense of long-term interruption as well as the cost of automation equipment are taken into account. In [4], from the metaheuristic algorithms, a genetic algorithm is selected in this paper to solve the optimal sectionalizing switches and reclosers placement problem. The dependability indices are optimized using this method, which takes into consideration the cost of automation equipment.

A long-term switch placement problem is presented in [5], considering load forecasting. The problem modeled as mixed-integer linear programming (MILP) formulation to include long-term load forecasting results obtained using the historical load data. A two-stage optimization algorithm is proposed in [6] to find the optimal number and place of automatic switches in the presented distribution network. In the first stage, the greedy optimization algorithm is applied to solve the restoration problem in the presence of contingencies. In the second stage, a cost function containing the ECOST and annual automatic switches cost is modelled and minimized due to the chosen locations of switches. An optimal distributed generation-based sectionalizing switches placement is investigated in [7] for distribution networks. The proposed problem is modeled as a mixed-integer non-linear programming (MINLP) problem and contains the cost of short-term outages and automation equipment. Optimal switch placement in the Ahvaz city distribution network is proposed in [8], in which the genetic algorithm is used to enhance the grid reliability. A model introduces in [9] to find the optimal number and location of automation devices. The presented model takes the benefit of MIP formulation making contributions to solving this extensive problem in an efficient run time and ensures the global optimum solution. The goal of this model is to minimize the total interruption and equipment costs.

Although valuable works have been done in the field of the optimal switch placement, but there are still many problems and deficiencies that need to be addressed properly. Briefly, the shortcomings of previous references are as follows:
a) The determination of the number and location of only one type of switch, mainly sectionalizes, is taken into account.
b) In the previously mentioned studies, switches are normally located in the predetermined positions in the feeders by default.
c) The calculation of operating cost function and the reliability service simultaneously in determining the best network automation solution is not addressed.

To address the shortcomings and drawbacks of previous literature, this paper attempts to provide the following contributions. The objective of this paper is to solve a multiobjective function in order to find compromised solutions both to enhance the reliability by optimal allocation of remote control switches (RCS) and minimize the total operating costs. In most of the prior works, where the optimal placement of switches has been considered, the number of RCS has been taken as fixed. In some literature where the number of RCS has been considered as a variable, multiobjective problem formulation has not been considered. In the present paper, both the number and location of RCS have been considered as variable, and a multiobjective function has been formulated. This paper develops a multiobjective model, wherein the primary objective, optimal RCS is implemented to minimize the total operating costs, while in the second objective, the reliability improvement is taken into account. A novel modified non dominated sorting genetic algorithm (MNSGA-II) is presented in this paper to reach the optimum global solutions. The MNSGA-II achieves solutions to the RCS placement problem instead of a single solution. This makes the proposed solution more adaptable to different utilities' circumstances and eases the decision-making. The following is a quick summary of the paper's novelties and contribution:

❖ Applying a robust optimization algorithm like MNSGA-II to solve the proposed complicated MINLP multiobjective problem.
❖ Applying the large-scale and practical distribution test systems like the fourth feeder of the TractorSazi substation post.
❖ Developing a mathematical methodology based on MINLP to solve the suggested optimal RCS placement problem for efficient reliability enhancement in distribution grids considering the cost of power losses, construction of maneuvering points, installation of RCSs, and maintenance and operation.

The backward-forward sweep-based method is explained in Section II. The mathematical formulation of the suggested problem is described in Section III. Section IV provides an overview of the MNSGA-II. Section V contains the simulation results. The conclusion is summarized in Section VI.

## II. BACKWARD-FORWARD SWEEP BASED ALGORITHM

The backward sweep and the forward sweep are two phases of this procedure. Backward sweep calculates voltage and current from the farthest bus from the beginning bus, applying KVL and KCL rules. The downstream voltage is determined starting from the beginning bus in the forward sweep. The bus-branch information is used as the algorithm's input data. The active and reactive power flows for sending and receiving buses, impedance, and susceptance of all branches are essential basic information. The major steps of the suggested algorithm are listed as follows: [10], [11].

1) Assume rated voltages at end nodes only for 1st iteration and equals the value computed in the forward sweep in the subsequent iteration.
2) Start with end node and compute the node current. Apply the KCL to determine the current flowing from node i-1 towards node i.
3) Compute with this current the voltage at *ith* node. Continue this step till the junction node is reached. At junction node the voltage computed is stored.
4) Start with another end node of the system and compute voltage and current as in step 2 and 3.
5) Compute with the most recent voltage at junction node.
6) Similarly, compute till the reference node.
7) Compare the calculated magnitude of the rated voltage at reference node with specified source voltage.

$$I_i^t = \left(\frac{P_i + jQ_i}{V_i^{t-1}}\right)^* \quad (1)$$

$$I_{i-1,i}^t = I_i^t + \sum_k I_{i,k}^t \quad k \geq i \quad (2)$$

$$V_i^t = V_{i-1}^{t-1} - (R_{i-1,i} + jX_{i-1,i})I_{i-1,i}^t \quad (3)$$

Stop if the voltage difference is less than specified criteria, otherwise forward sweep begins.
*Forward Sweep:*
1) Start with reference node at rated voltage.

2) Compute the node voltage in forward direction from reference node to end nodes.
3) Again, start backward sweep with updated bus voltage calculated in forward sweep. After calculating node voltages and line currents using standard backward-forward sweep algorithm, the line losses are calculated.

### III. Problem Formulation

The suggested method applies a multi-objective mathematical programming paradigm with two competing goals: 1) minimizing the cost of RCS deployment and 2) improving the reliability of service. The primary objective is to keep the cost of sectionalizing RCSs as low as possible. Various reliability indices can be used to construct the second objective. To measure service reliability, the energy not supplied is used.

*First objective: RCS placement cost function:* The following costs are involved in maneuvering devices placement and maneuvering points design in distribution networks: i) Damages caused by not supplying energy to customers; ii) Costs related to the maneuvering devices construction, including price and installation cost; iii) Costs related to the maneuvering points, including switch and line price, and iv) Annual cost of equipment maintenance and operation. In this case, the goal is to minimize the sum of these costs.

$$\text{Min } F_1 = \begin{cases} \sum_{i=1}^{n} DS_i \times CS + \sum_{j=1}^{m} DT_j \times CT_j \\ + \sum_{t=1}^{ny} P_W^t \times MC + \sum_{l=1}^{L} P_{Loss}^t \times K_{Loss} \end{cases} \quad (4)$$

$$P_W = \frac{1+Infr}{1+Intr} \quad (5)$$

*Second objective: Energy not supply function:* In this paper, reliability studies based on the analytical method are performed. This analytical method is to evaluate reliability based on failure mode and effect analysis. At first, the failure modes that affect each load point (distribution posts) are identified. Then, by evaluating their effect, reliability index (ENS) is calculated in each load point. In this calculation, the impact of network configuration, the existence of sectionalizing RCSs and their locations for fault separation, and the possibility of supplying load points through support communication lines are considered. The amount of ENS related to domestic and industrial consumers plays a very important role in distribution system automation. Equation (6) shows the consequences of energy not supply function in distribution networks.

$$\text{Min } F_2 = \sum_{i=1}^{nLP} ENS_T = \sum_{i=1}^{nLP} IC_i \times ENS_i \times K_i \quad (6)$$

$$IC_i = \begin{cases} AIC_i(\text{res}).P_i(\text{res}) + AIC_i(\text{com}).P_i(\text{com}) \\ + AIC_i(\text{ind}).P_i(\text{ind}) + AIC_i(\text{agr}).P_i(\text{agr}) \\ + AIC_i(\text{gen}).P_i(\text{gen}) \end{cases} \quad (7)$$

$$\lambda_s = \sum_{i \in A} \lambda_i \quad (\text{f / yr}) \quad (8)$$

$$U_s = \sum_{i \in A} \lambda_i r_i \quad (\text{h / yr}) \quad (9)$$

$$r_s = \frac{U_s}{\lambda_s} \quad (\text{h / f}) \quad (10)$$

$$ENS = PU_s \quad (\text{kWh / yr}) \quad (11)$$

The average failure rate and average outage time are given in equations (8) and (9), respectively. Also, the annual average outage time and average ENS are presented as equations (10) and (11), respectively.

The stochastic behavior of load demands is modeled by normal PDF, which is demonstrated in Eq. (12) [12].

$$f_L(a) = \frac{1}{\sqrt{2\pi} \times \sigma} \exp\left(\frac{(a-\mu^2)}{2 \times \sigma^2}\right) \quad (12)$$

### IV. Providing an Overview on MNSGA-II

Non-dominated sorting for fitness tasks is utilized by the MNSGA-II [13]. Both polynomial mutation and binary crossover produce different generations, and event choice is later employed to pick out the populace for the subsequent generation. Elitism resolves the trouble of dropping true answers for the duration of the optimization cycle because of chance impacts. The MNSGA-II with dynamic crowding distance (DCD) is utilized to tackle the issue [14].

*1) Simulated binary crossover (SBC)*

In common, SBC places the strain on producing generation close to the parents. This crossover ensures that the volume of the youngsters or generation is equal to the parents' volume and additionally supports that close to determine people are monotonically much more likely to be selected as youngsters than children remote from the parents with inside the answer space [13]. The process for detecting the generation answers $x_i^{(1,t+1)}$ and $x_i^{(2,t+1)}$ from parent answers $x_i^{(1,t)}$ and $x_i^{(2,t)}$ is presented as follow. $u_i$ is a random number that is define between 0 and 1. Afterwards, by using probability distribution function, the $\beta_{qi}$ is obtained as equation (13):

$$\beta_{qi} = \begin{cases} (2u_i)^{1/\eta_c+1} & , \text{if } u_i \leq 0.5; \\ (\frac{1}{2(1-u_i)})^{1/\eta_c+1} & , \text{otherwise.} \end{cases} \quad (13)$$

In equation (13), the indicator $\eta_c$ is a positive number. Since finding $\beta_{qi}$, the children answers are computed as equations (14) and (15):

$$x_i^{(1,t+1)} = 0.5\left[(1+\beta_{qi})x_i^{(1,t)} + (1-\beta_{qi})x_i^{(2,t)}\right] \quad (14)$$

$$x_i^{(1,t+1)} = 0.5\left[(1-\beta_{qi})x_i^{(1,t)} + (1+\beta_{qi})x_i^{(2,t)}\right] \quad (15)$$

*2) Polynomial mutation*

The possibility of making an answer close to the parents is better than the possibility of making one remote from it. The form of the probability distribution is immediately managed via an outside parameter and stayed constant at some stage in the iterations. Like within side the SBC operator, the probability distribution also can be a polynomial function, in place of a normal distribution:

$$y_i^{(1,t+1)} = x_i^{(1,t+1)} + (x_i^{(U)} - x_i^{(L)})\delta_i \quad (16)$$

Parameter $\delta$ is computed from the polynomial probability distribution

$$P(\delta_i) = 0.5(\eta_m + 1)(1 - |\delta|)^{\eta_m} \quad (17)$$

$$\delta_i = \begin{cases} (2r_i)^{1/\eta_m + 1} - 1, & \text{if } r_i < 0.5 \\ 1 - [2(1-r_i)]^{1/\eta_m + 1}, & \text{if } r_i \geq 0.5 \end{cases} \quad (18)$$

For handling the bounded decision variables, the mutation operator is modified for two regions: $[x_i^{(L)}, x_i]$ and $[x_i, x_i^{(U)}]$.

*3) Dynamic crowding distance (DCD)*

In multi-objective evolutionary algorithms, the horizontal diversity of pareto front is very important. Horizontal diversity is often realized by removing excess individuals in the non-dominated set (NDS) when the number of non-dominated solutions exceeds population size. MNSGA-II uses crowding distance (CD) measure as given in Eq. (19) to remove excess individuals.

$$CD_i = \frac{1}{r}\sum_{k=1}^{r}\left|f_{i+1}^k - f_{i-1}^k\right| \quad (19)$$

The main shortcoming of CD is the lack of uniform variety in the occupied non-dominated answers. To overcome this issue, the dynamic crowding distance (DCD) approach is newly proposed in [16]. In this method, one individual with the minimum DCD amount is deleted every time and DCD is re-computed for the remaining individuals. The individuals DCD are computed as equation (20):

$$DCD_i = \frac{CD}{\log(1/V_i)} \quad (20)$$

where $CD_i$ is computed by Eq. (19) and $V_i$ is based on Eq. (21),

$$V_i = \frac{1}{r}\sum_{k=1}^{r}\left(\left|f_{i+1}^k - f_{i-1}^k\right| - CD_i\right)^2 \quad (21)$$

$V_i$ is the variation of CDs of individuals which are neighbors of the *i*th individual.

## V. SIMULATION RESULTS AND DISCUSSION

The suggested algorithm is used in a more complicated and effective system to investigate the applicability of the model in real conditions. The proposed test systems are the third feeder of the Azar substation with a distributed generation unit and first and third feeders of ElGoli substation which form a double feed feeder given in Figures 1 and 2, respectively.

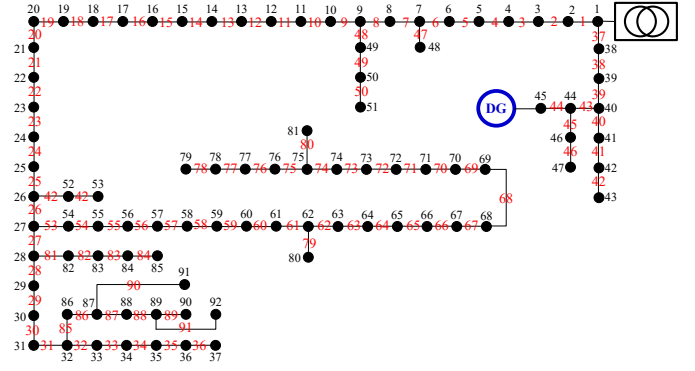

Fig. 1. Third feeder of Azar substation post

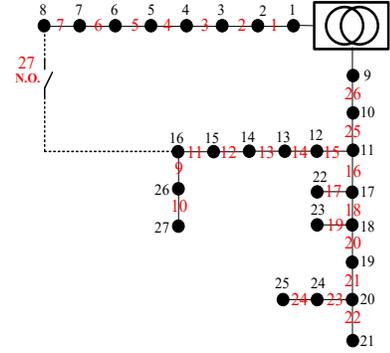

Fig. 2. First and third feeders of ElGoli substation post

1) The failure rate of all RCSs is assumed to be zero.
2) The cables in the test system are assumed to be of both overhead and underground types.
3) The minimum and maximum voltage limits are assumed between 0.95 and 1.05 pu, respectively [17].
4) The price of each switch is supposed to be $ 4,700. The maintenance cost is 2% of the investment cost. [18].
5) Repair time and failure rate are presented in Table I [19].

**Table I.** failure rate and repair rate of the system equipment

| Components | Failure rate ($\lambda$) | Repair time (r) |
|---|---|---|
| Transformers | 0.004 (f/yr) | 4 hr. |
| Distribution lines | 0.0075 (f/yr/km) | 2 hr. |

This article summarizes the steps, data collection, and implementation of simulation as follows:

**Step 1:** Obtaining static data from GIS containing the structure, load points, load data, and type of transformers.
**Step 2:** Enter the structure and data from GIS to MATLAB.
**Step 3:** Calculates the current flow of lines and voltage drop to implement the backward-forward power flow method.
**Step 4:** Find best optimum global solutions using MNSGA-II.
**Step 5:** Print the results of minimum ENS and operating costs.

Fig. 3 shows the amount of each section ENS relate to the third feeder of Azar. The ENS of each section is calculated by using the objective function (6) and the MNSGA-II is used to minimize the total amount of ENS to enhance the reliability service goals in the proposed distribution test systems. Figs. 4 shows the voltage profile after and before applying the proposed method and DG. As can be seen from this figure; after using the proposed model the voltage magnitude of each node is significantly smooth in comparison to the initial case. Moreover, Fig. 5 depicts the active power losses. Employing the proposed multi-objective approach can remarkably reduce the amount of power losses related to each line.

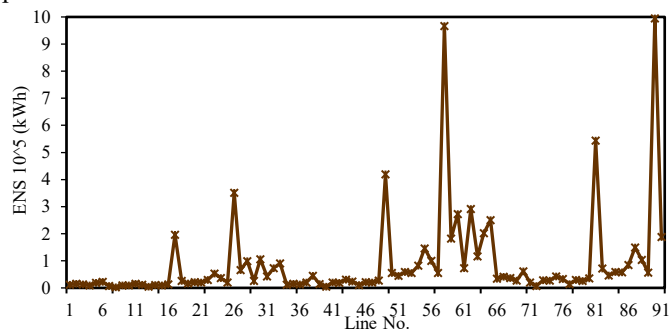
Fig.3. ENS related to the third feeder of the Azar substation.

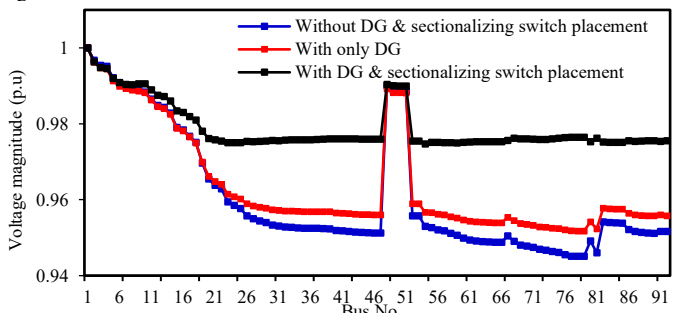
Fig. 4. Voltage profile of third feeder of the Azar substation.

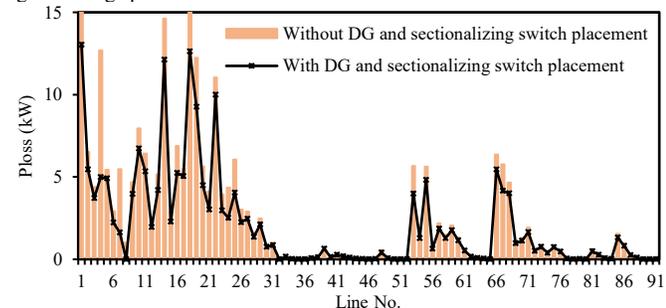
Fig. 5. Active power losses of the third feeder of the Azar substation.

Figure 6 shows the amount of each section ENS relate to the first and third feeders of ElGoli substation post. The ENS of each section is calculated by using the objective function (6) and the MNSGA-II is used to minimize the total amount of ENS to enhance the reliability service goals in the proposed distribution test systems. Fig. 7 shows the voltage profile after and before applying the proposed method. As can be seen from this figure, after using the proposed model the voltage magnitude of each node is significantly smooth in comparison to the initial case. Moreover, Fig. 8 depicts active power losses. Employing the proposed multi-objective approach can remarkably reduce the amount of power losses related to each line. Finally, these results deduced that optimal placement of remotely controlled automation devices has a good impact on the system technical specification.

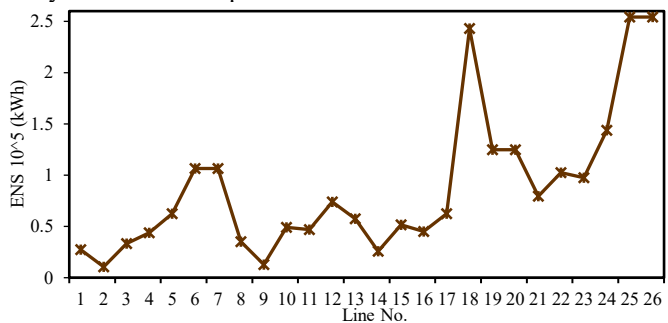
Fig. 6. ENS related to the first and third feeders of ElGoli substation.

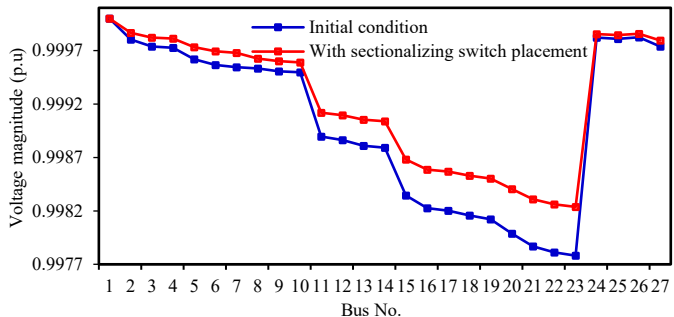
Fig. 7. Voltage profile first and third feeders of ElGoli substation.

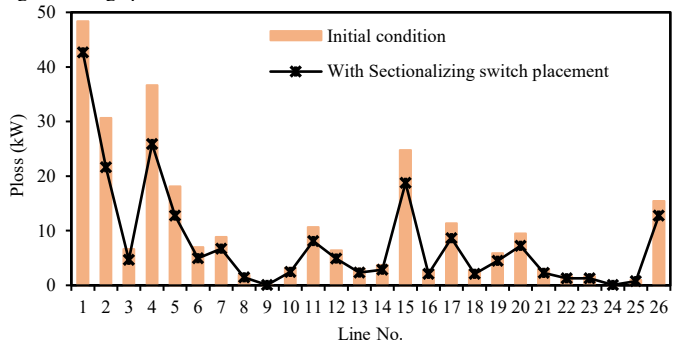
Fig. 8. Active power losses of the first and third feeders of ElGoli substation.

Tables II and III show the comparison of the real and optimal placement of RCSs in the proposed distribution system. These tables declare that using a good performance with a high convergence rate algorithm like MNSGA-II has a major impact on the obtained results. So that, in Table IV and Table V the minimum cost of switching equipment in optimal placement condition is reduced near to %79 and %88.5 of its experimental placement condition, respectively. The number of RCSs is reduced from 5 to 4 and the proposed algorithm finds the critical points of the system according to the first objective function for locating these RCSs. Finally, in Tables II and III, the total ENS of all sections is reduced near to %18 and %7, respectively, which proves using the MNSGA-II algorithm improves the reliability service of the distribution system.

**Table II.** Comparison of the real and optimal placement of RCSs in the proposed distribution network

| Item | Real and experimental placement condition | Optimal placement condition using MNSGA-II |
|---|---|---|
| Minimum cost of switching equipment | 260907.77 ($/yr) | 229945.7308 ($/yr) |
| Number of sectionalizing RCSs | 6 | 5 |
| Switch placement | S21, S37, S48, S73, S79, S85 | S22, S38, S69, S81, S85 |
| Total ENS | 5510870.878918 (kWh) | 4942485.0932 (kWh) |

Table III. Comparison of the real and optimal placement of RCSs in the first and third feeders of ElGoli substation

| Item | Real and experimental placement condition | Optimal placement condition using MNSGA-II |
|---|---|---|
| Minimum cost of switching equipment | 10097.5390 ($/yr) | 8935.8753 ($/yr) |
| Number of sectionalizing RCSs | 1 | 1 |
| Switch placement | S27 | S12 |
| Total ENS | 2794006.4880 (kWh) | 2611221.0176 (kWh) |

Figs. 9 and 10 display the optimal Pareto curve obtained from MNSGA-II. All dominated members are eliminated, and non dominated members remain. The optimal solution is the blue square around it. This point is the best and optimum global solution, which is the minimum amount of operating cost and average ENS of all sections have been achieved.

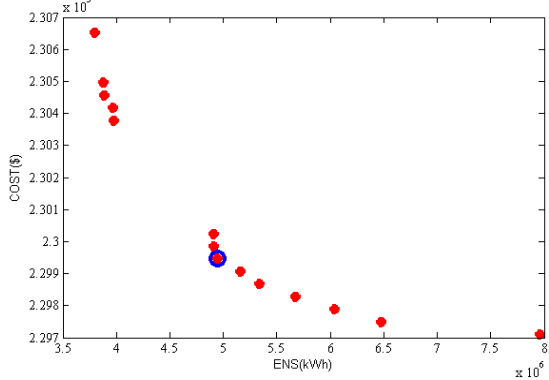

Fig. 9. Pareto curve of third feeder of the Azar substation.

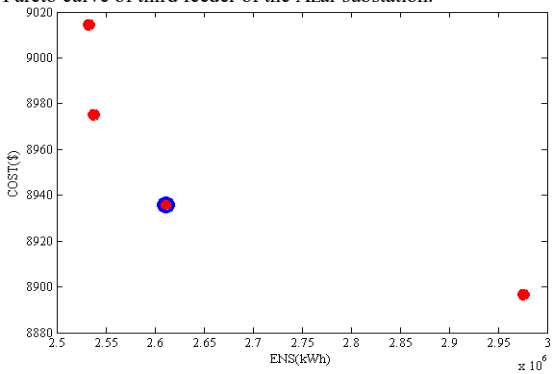

Fig. 10. Pareto curve of first and third feeders of ElGoli substation.

## VI. CONCLUSION

This paper has presented a novel multi-objective model that incorporates two objectives: minimization of cost of sectionalizing RCSs and improvement of reliability indices. The presented results indicate that the system reliability and cost are affected by the number and cost of RCSs, their locations, and the availability of alternative supplies. The proposed approach has been applied to an actual distribution networks, and the effectiveness of the MNSGA-II-based method to solve this MINLP problem has been demonstrated through the results obtained. The proposed work minimizes the cost of switching equipment to %79 and %88.5 of its real and experimental condition in third feeder of Azar and first and third feeder of Elgoli substation posts, respecetevely. Also, some specifications of the distribution test system like voltage profile and total active power loss have been improved by applying the suggested approach.

## REFERENCES

[1] Bagheri, M. Bagheri, and A. Lorestani, "Optimal reconfiguration and DG integration in distribution networks considering switching actions costs using tabu search algorithm," *Journal of Ambient Intelligence and Humanized Computing*, vol. 12, no. 7, pp. 7837-7856, 2021.

[2] A. Shahbazian, A. Fereidunian, and S. D. Manshadi, "Optimal switch placement in distribution systems: A high-accuracy MILP formulation," *IEEE Transactions on Smart Grid*, vol. 11, no. 6, pp. 5009-5018, 2020.

[3] M. Elkadeem, M. Alaam, and A. M. Azmy, "Optimal automation level for reliability improvement and self-healing MV distribution networks*,"* in *2016 Eighteenth International Middle East Power Systems Conference (MEPCON)*, 2016, pp. 206-213: IEEE.

[4] A. Abdolahi, R. Ajabi Farshbaf, A. Abbaspour, and M. Khodayari, "Reliability enhancement by strategic placement of remotely-controlled automation devices: A case study on Tabriz distribution network," *Res. Technol. Electr. Ind.*, vol. 1, no. 1, pp. 36–45, Mar. 2022.

[5] J. Forcan and M. Forcan, "Optimal placement of remote-controlled switches in distribution networks considering load forecasting," *Sustainable Energy, Grids and Networks,* vol. 30, p. 100600, 2022.

[6] M. Isapour Chehardeh and C. J. Hatziadoniu, "Optimal placement of remote-controlled switches in distribution networks in the presence of distributed generators," *Energies,* vol. 12, p. 1025, 2019.

[7] A. Heidari, V. G. Agelidis, M. Kia, J. Pou, J. Aghaei, M. Shafie-Khah*, et al.*, "Reliability optimization of automated distribution networks with probability customer interruption cost model in the presence of DG units," *IEEE Transactions on Smart Grid,* vol. 8, pp. 305-315, 2016.

[8] H. Karimi, T. Niknam, J. Aghaei, M. Ghasemi Garpachi, and M. Dehghani, "Switches optimal placement of automated distribution networks with probability customer interruption cost model: A case study," *International Journal of Electrical Power & Energy Systems,* vol. 127, p. 106708, 2021.

[9] M. Farajollahi, M. Fotuhi-Firuzabad, and A. Safdarian, "Simultaneous placement of fault indicator and sectionalizing switch in distribution networks," *IEEE Transactions on Smart Grid,* vol. 10, pp. 2278-2287, 2018.

[10] N. Taghizadegan, S. Cheshmeh Khavar, A. Abdolahi, F. Arasteh, and R. Ghoreyshi, "Dominated GSO algorithm for optimal scheduling of renewable microgrids with penetration of electric vehicles and energy storages considering DRP," *Int. J. Ambient Energy*, vol. 43, no. 1, pp. 6380–6391, Dec. 2022.

[11] M. Amini, A. Jalilian, and M. R. P. Behbahani, "Fast network reconfiguration in harmonic polluted distribution network based on developed backward/forward sweep harmonic load flow," *Electric Power Systems Research*, vol. 168, pp. 295-304, 2019.

[12] N. T. Kalantari, A. Abdolahi, S. H. Mousavi, S. C. Khavar, and F. S. Gazijahani, "Strategic decision making of energy storage owned virtual power plant in day-ahead and intra-day markets," *J. Energy Storage*, vol. 73, p. 108839, Dec. 2023.

[13] F. Shabani-Naeeni and R. G. Yaghin, "Integrating data visibility decision in a multi-objective procurement transport planning under risk: A modified NSGA-II," *Applied Soft Computing*, vol. 107, p. 107406, 2021.

[14] L. Pan, W. Xu, L. Li, C. He, and R. Cheng, "Adaptive simulated binary crossover for rotated multi-objective optimization," *Swarm and Evolutionary Computation*, vol. 60, p. 100759, 2021.

[15] K. Deb, "Multi-objective optimisation using evolutionary algorithms: an introduction," in *Multi-objective evolutionary optimisation for product design and manufacturing*: Springer, 2011, pp. 3-34.

[16] B. Luo, J. Zheng, J. Xie, and J. Wu, "Dynamic crowding distance? A new diversity maintenance strategy for MOEAs," in *2008 Fourth International Conference on Natural Computation*, 2008, vol. 1, pp. 580-585: IEEE.

[17] S. Cheshme-Khavar, A. Abdolahi, F. S. Gazijahani, N. T. Kalantari, and J. M. Guerrero, "Short-term scheduling of cryogenic energy storage systems in microgrids considering CHP-thermal-heat-only units and plug-in electric vehicles," *J. Oper. Autom. Power Eng.*, vol. 12, no. 3, pp. 233–244, Aug. 2024.

[18] Y. Tan et al., "Repulsive firefly algorithm-based optimal switching device placement in power distribution systems," *Global Energy Interconnection*, vol. 2, no. 6, pp. 489-495, 2019.

[19] A. Abdolahi, N. Taghizadegan, M. R. Banaei, and J. Salehi, "A reliability-based optimal μ-PMU placement scheme for efficient observability enhancement of smart distribution grids under various contingencies," *IET Science, Measurement & Technology*, 2021.